\documentclass{article}
\newcommand{\bfr}{\begin{flushright}}
\newcommand{\efr}{\end{flushright}}
 
\begin{document}
\title{A New Vector-Tensor Theory and Higher-Dimensional Cosmology
}
\author{Katsuhiko Yoshida\\
Department of Physics,
Tokyo Metropolitan University,\\ Setagaya, Tokyo 158, Japan\\
Kiyoshi Shiraishi\\
Institute for Nuclear Study, University of Tokyo,\\ Midori-cho, Tanashi,
Tokyo 188, Japan
}
\date{Physica Scripta {\bf 43}, No.~2, pp.~129--132 (1991)
}
\maketitle
\begin{abstract}
A new vector-tensor model of classical gravity, which contains
coupling between the field strength of the vector field and the
curvature tensors in six dimensions, is proposed.
Cosmological solutions of the scale factors in this model with
the compactified space are obtained. The generalization of the
model is also considered.
\end{abstract}

\section{INTRODUCTION}

The theory of general relativity seems the most probable
theory which describes a classical behavior of space-time
geometry at a large scale. To seek for the other possibilities
in the description of gravitation, various modifications of
Einstein's gravity have been investigated by many authors in the
past several decades. In some of the theories gravity is coupled
to other fields; for example, the well-known Brans-Dicke theory~\cite{1}
has a crucial coupling between the scalar curvature of the
space-time manifold and a scalar field. On the other hand, we
are able to consider a theory which is described by an action
that includes higher-derivative terms of the metric tensor, such
as a curvature-squared term~\cite{2}. Moreover, from a relation to
unified theories, gravity can be derived by an action which is
not the pure Einstein--Hilbert action.  These theories can be
formulated in more than four-dimensions, such as Kaluza--Klein 
theory \cite{3} and the string theory \cite{4}. We must note that
multidimensional theories need not involve the gravity in
higher dimensions, since they have only to induce a theory
which is consistent with Einstein's general relativity in four
dimensions \cite{5}. True theory of gravity may have some of
the above-mentioned structures, and then it is worth studying
possible combinations of the ideas about the gravitation.

As another difficulty in the theory of gravity, there are
many problems in a quantization of gravity; one of them is that
renormalization has been impossible because the coupling constant
of gravitons, i.e., Newton's constant $G$, has dimensions.
Although we will not consider the quantization of gravity in this
letter, we will construct a model as a theory with a dimensionless
coupling in higher dimensional space-time. And the
theory induces a correct dimensionality of the effective coupling
of gravity in four dimensions.

We start with some possible hypotheses.
First, a coupling constant in our model is dimensionless. Second,
there are higher-derivative terms of the metric but an action
has first
order of second derivatives of the metric. In this case,
evolution equations can be obtained from the action as in the
case with the Eintein--Hilbert action, and we also expect that 
the solution of the equation of motion could not have so many branches
(cf.~ref.~\cite{6}).

Furthermore, we require that our model can have more than
four dimensions, and then it does not describe Einstein gravity
in the higher dimensions (cf.~ref.~\cite{5}). Of course, we require
that our model consistently describe Einstein gravity after
compactification of an extra space.

In this letter, we will show that our model has a good
symmetry and weIl-behaved cosmological solution of homogeneous
spaces.

To construct an action which satisfies the requirements
above, we choose an action which is linear in curvature tensors.
Further, we introduce a vector field coupled to the curvature
nonminimally.

We deal with our vector-tensor model in six dimensions. The
generalization to the  other dimensions will be discussed
later. Our action is analogous to four-dimensional 
Euler form \cite{6,7}.
That is:
\begin{equation}
I=-\frac{\gamma}{16}\int
d^6x\sqrt{-g}\delta^{ABCD}_{EFGH}F_{AB}F^{EF}R_{CD}{}
^{GH}+(\mbox{surface terms})\,,
\label{eq1}
\end{equation}
where the generalized Kronecker's symbol is defined by
\begin{equation}
\delta^{ABCD}_{EFGH}=4!\delta_{[E}^{A}\delta_{F}^{B}
\delta_{G}^{C}\delta_{H]}^{D}\,.
\label{eq2}
\end{equation}
Here $\gamma$ is a dimensionless coupling because our action is
defined in six dimensions. Horndeski showed \cite{8} that this form of
the action in four dimensions leads to the conservation of the
``electric charge'' as well as the energy-momentum. The surface
terms can be added to remove the second-derivative of the metric,
as is the Gibbons--Hawking surface term in Einsteinean gravity \cite{9}.
We consider two dimensions compactified to $S^2$, and we put
an ansatz on the field strength of the vector field on $S^2$ as:
\cite{10}
\begin{equation}
F_{ab}=q \varepsilon_{ab}\,,
\label{eq3}
\end{equation}
where $q$ is a constant. This ansatz is used throughout this
letter. This satisfies the equation of motion extracted from the
action (\ref{eq1}), at least when the background metric has a
homogeneous structure as the case we will consider in this letter.

If we want to include the matter, we add the action of the matter
fields in the form of the six-dimensional integral.

In order to show several cosmological solutions to the field equations
obtained from our action, we will use the metric of the form:
\begin{eqnarray}
ds^2&=&-N^2dt^2+a^2(t)\left(1+\frac{k}{4}r^2\right)^{-2}
\{dr^2+r^2(d\theta^2+\sin^2\theta
d\phi^2)\}\nonumber \\
& &+b^2(t)(d\omega^2+\sin^2\omega d\chi^2)\,.
\label{eq4}
\end{eqnarray}

By using this metric and the ``monopole'' ansatz (\ref{eq3}) for the
vector field, the action reduces to:
\begin{equation}
I=3\gamma q^2\int
dt\left\{\frac{1}{N}\left[\frac{a\dot{a}^2}{b^2}-2
\frac{a^2\dot{a}\dot{b}}{b^3}\right]-Nk\frac{a}{b^2}\right\}\,.
\label{eq5}
\end{equation}
The field equations can be obtained by variation of the action
(\ref{eq5}) with respect to the functions $a, b$ and $N$: the variation
with respect to $N$ leads to the analogous equation to the
time-time component of Einstein equations after taking $N\rightarrow
1$.  In the presence of the matter Lagrangian, the field equations are:
\begin{eqnarray}
\left(\frac{\dot{a}}{a}\right)^2+2\frac{\ddot{a}}{a}+2\frac{\dot{a}}{a}
\frac{\dot{B}}{B}+\frac{\ddot{B}}{B}+\frac{k}{a^2}&=&-\frac{1}{\gamma
q^2B^2}(p)_{matter}\,,
\label{eq6a}
\\
\left(\frac{\dot{a}}{a}\right)^2+\frac{\ddot{a}}{a}
+\frac{k}{a^2}&=&\frac{1}{3\gamma
q^2B^2}(p')_{matter}\,,
\label{eq6b}
\\
\left(\frac{\dot{a}}{a}\right)^2+\frac{\dot{a}}{a}\frac{\dot{B}}{B}
+\frac{k}{a^2}&=&\frac{1}{3\gamma
q^2B^2}(\rho)_{matter}\,,
\label{eq6c}
\end{eqnarray}
where $B\equiv b^{-2}$, and $(\rho, p, p')_{matter}$ represent
(the energy density, pressure in the three-dimensional space, pressure
in the extra two-dimensional space) respectively.

At first in order to inspect whether our model has a well-behaved
cosmological solution, we solve a vacuum solution as the simplest case.
The field equations can be solved exactly,
\begin{equation}
a(t)=(c_1t-kt^2)^{1/2}\,,\quad (c_1: \mbox{const.})
\label{eq7a}
\end{equation}
and
\begin{equation}
b(t)=c_2\left|\frac{t(c_1-kt)}{(c_1-2kt)^2)}\right|^{1/4}\,,
\quad (c_2: \mbox{const.})\,.
\label{eq7b}
\end{equation}

Although the radius of the compactified inner space diverges at
$t=c_1/2$, when $k=1$, we can consider that we have not reached the
time yet; the early evolution in this vacuum solution is relevant for
the present universe. The behavior of the scale factor of our space
($a$) corresponds to a solution with radiation in the four-dimensional
Friedmann-Robertson-Walker (FRW) model. The
behavior of $a$ is determinded by eq.~(\ref{eq6b}).

Next we consider solutions in the presence of matters. Here
we deal with some typical examples of matters, such as the quadratic of
the field strength, dust matter, four-dimentional radiation and
six-dimensional radiation. Some of the field equations with matter can
be exactly solved and we first treat this kind of solutions.

We will show a solution with an extra contribution of the field
strength. Namely, we add the usual Maxwell-type action of the quadratic
term in the field strength to our action.
That is to say, we incorporate the term
$\int d^6x\sqrt{-g}(\beta/4)F_{MN}F^{MN}$ into the action. We substitute
the ``monopole'' form of the field strength in the extra space as the
previous case.  The equation of motion of the scale factor of
three-dimensional space is solvable, and the solution is given, in the
particular case where
$k=0$, as:
\begin{equation}
a(t)=[c_1 \sinh(2\alpha')^{1/2}t]^{1/2}\,,\quad (c_1: \mbox{const.})\,,
\label{eq8}
\end{equation}
where $\alpha'\equiv\beta/6\gamma$.

In this case, the scale of the extra space is:
\begin{equation}
b(t)=c_2\left|\frac{\tanh(2\alpha')^{1/2}t}{\cosh(2\alpha')^{1/2}t}
\right|^{1/4}\,,\quad(c_2:
\mbox{const.})\,.
\label{eq9}
\end{equation}
The field equations can be solved exactly even if $k\ne 0$, after some
manupilations. But we omit the exhibition of the solutions.

Just as the vacuum solution, the behavior of the scale factor of our
space corresponds to the solution in the presence of
radiation as well as a cosmological constant in four-dimensional
FRW model.

Next we consider dust and four-dimensions radiation as a
matter. which are analogue of matter and radiation dominated era
in four-dimensional FRW model. The field equation of the
three-dimensional scale factor is the same as a vacuum case:
\begin{equation}
a(t)=(c_1t-kt^2)^{1/2}\,.
\label{eq10}
\end{equation}
And the scale factor of the compactified extra space in each case is
\begin{eqnarray}
& &\mbox{For dust matter:}\nonumber \\
& &b_d=\left[\frac{C'}{3k\gamma
q^2(c_1t-kt^2)^{1/2}}+c_2\frac{c_1-2kt}{(c_1t-kt^2)^{1/2}}\right]^{-1/2}\,,
\quad (k\ne 0)\\ 
& &b_d=\left[\frac{2C'}{3\gamma
q^2c_1^{3/2}}t^{1/2}+c_2\frac{c_1^{1/2}}{t^{1/2}}\right]^{-1/2}\,, 
\quad (k=0)\\
& &\mbox{For four-dimensional radiation:}\nonumber \\
& &b_r=\left[\frac{4C}{3\gamma
q^2}+c_2\frac{c_1-2kt}{(c_1t-kt^2)^{1/2}}\right]^{-1/2}\,,
\label{eq11b}
\end{eqnarray}
where $b_d$ and $b_r$ are the solution in the presence of a dust matter
and the one in the presence of four-dimensional radiation,
respectively. $c_1$ and $c_2$ are integration constants. Here the
amount of the matter is represented as $\rho_{dust}=C'/(a^3b^2)$ and
$\rho_{4\mbox{-}rad.}=C/(a^4b^2)$ in each case.

Finally we show the solution of the universe dominated by
six-dimensional radiation. Since in this case the field
equations cannot be solved in the analytic form, we show the
leading behaviors of the scale factors at the early stage of the
evolution. For six-dimensional radiation, its energy density is
written as $\rho=\rho_0(a^3b^2)^{-6/5}$ ($\rho_0=$const.).

Here we take $k=0$, because we need a solution which
corresponds to the early universe. We assume that solutions of
the scale factors in this case are proportional to the power of
cosmic time, thus we can write $a=t^\alpha$ and $b=t^\beta$,
respectively. Substituting these in
eqs.(\ref{eq6a},\ref{eq6b},\ref{eq6c}), we find a set of solutions for
$\alpha$ and $\beta$:
\begin{equation}
\alpha=\frac{5}{9}\quad \mbox{and} \quad \beta=0\,.
\label{eq12}
\end{equation}
These behaviors of the scale factors are very close to the vacuum case
($\alpha=1/2$). And this fact indicates that more moderate change of
the scale of the extra space is permitted in the presence of the
six-dimensional rediation.

So far we obtain some cosmological solutions of our six-dimensional
vector-tensor model including the exact solutions. For the vacuum
solution and the solution in the presence of the extra Maxwell term,
there are relations to other cosmological models in other dimensions.

Changing variables in our model as $1/b^2=B$ reduces our action
(\ref{eq1}) to the five-dimensional Kaluza-Klein action with taking the
metric, $ds^2=-dt^2+a^2d\Omega^2+b^2dy^2$. Accordingly, the vacuum
solution in our model corresponds to the vacuum solution in the
flve-dimensional model, up to the substitution of $1/b^2=B$. In the same
way, the Maxwell term $\beta F^2/4$, which is equivalent to $\beta
(2q^2)/4b^4$ by use of the monopole ansatz, becomes the form of
$\Lambda'B^2$ by changing variable as $1/b^2=B$ and $\Lambda'=\beta
q^2/2$. This term corresponds to the cosmological term in that of
five-dimensional Kaluza-Klein theory. It is known \cite{11} that the
five-dimensional Kaluza-Klein cosmological solution with a cosmological
constant (including vanishing cosmological constant) can be related to
the universe dominated by four dimensional radiation with an
appropriately normalized cosmological constant.

As an extension of our vector-tensor model, we would like to consider a
model with more than six dimensions, that is, $4+N$ dimensions. A
natural extension can be discovered in $(4+N)$ dimensions is described
by an action which contains an antisymmetric tensor field. The possible
action is
\begin{eqnarray}
&I&=-\frac{\gamma_N}{4(N!)^2}\int d^{4+N}x\sqrt{-g}\delta^{A_1\cdots
A_{N+2}}_{B_1\cdots B_{N+2}}H_{A_1\cdots
A_{N}}H^{B_1\cdots
B_{N}}R^{B_{N+1}B_{N+2}}{}_{A_{N+1}A_{N+2}}\nonumber \\
& &+(\mbox{surface terms})\,,
\label{eq13}
\end{eqnarray}
where Kronecker's delta is generalized to have $2(N+2)$ indices.
Here $\gamma_N$ is not dimensionless, regretfully.
But the dimension of $\gamma_N$ is $(\mbox{mass})^{2-N}$; then it is not
so bad in the sense of powercounting, though we never consider the
quantization here. If we assume the ``expectation value'' of the form 
$H_{\dots}=q
\varepsilon_{\dots}$ on the extra $N$-manifold, the vacuum solution in
this model can be solved exactly.  At early stage of the evolution of
our three-space behaves like $t^{1/2}$, whereas the scale factor of the
extra space behaves like
\begin{equation}
b(t)\approx t^{\frac{1}{2N}}\,.
\label{eq14}
\end{equation}
Therefore, if $N$ is sufficiently large, time evolution of the
coupling ``constant'' like the gravitational coupling and gauge
couplings (if we add), which are concerned with $b$, can be so
small as to be consistent with the observations \cite{12}.

In summary, we obtain the cosmological solutions in our vector-tensor
model in six dimensions with compactification in the presence of 
various matter fields. The relation to the other
dimensional cosmological model makes the derivation of the
solutions simple. Note that the extra space in our model admits
a non-Abelian isometry.  The generalization of our model to
higher-dimensions is straightforward, and the rapid change in
the coupling constant can be avoided if we take the model with
sufficiently large number of total dimensions.

Spherically symmetric vacuum solution can be obtained
analytically because of the correspondence with five-dimensional
``black hole'' solution by the similar substitution as in this
letter; in particular, Schwarzschild solution with extra space of
constant radius is permitted. But if we consider ``charged''
object, the solution in our vector-tensor model differs from the
Reissner-Nordstrom solution in general relativity. The numerical
study on the solution of this type is in progress and will be
reported.

Another interesting topic is the making of the whole universe
as a topological defect \cite{13}. We start with flat six
dimensions. We introduce a complex scalar minimally coupled to the
vector field in order to construct a Nielsen-Olesen vortex \cite{14}. 
Since the field strength is concentrated at the core of the vortex, the
gravity ``exists'' only there. Now we obtain a ``thin''
universe with spatial three dimensions. Further study about this
scenario including examination of the ``confinement'' in the universe
and the back reaction is a future problem.

We have treated our models as a classical theory.  Although
the coupling constants have favorable dimensionality, quantization
of our models is not straightforward. This is typically shown by
the fact that the kinetic term of the vector field has ``wrong''
sign in the compactified background. We think that this is not a
great drawback, because the component with negative sign always
exists in the Einstein gravity and the minimal corrections of the
theory. We must study the quantization of our model as well as
the gravity theories in general.

\section*{Acknowledgements}
The authors would like to thank A.~Nakamula for some comments.

This work is supported in part by the Grant-in-Aid for Encouragement of
Young Scientist from the Ministry of Education, Science and Culture (\#
63790150).

One of the authors (KS) is grateful to the Japan Society for the
Promotion of Science for the fellowship. He also thanks Iwanami
F\=ujukai for financial aid.


\end{document}